# Anonymizing Periodical Releases of SRS Data by Fusing Differential Privacy

Yi-Yuang Wu
*Dept. of Comp. Sci. and Info. Eng.*
*National University of Kaohsiung*
Kaohsiung, Taiwan
m1085507@mail.nuk.edu.tw

Zhi-Xun Shen
*Dept. of Comp. Sci. and Info. Eng.*
*National University of Kaohsiung*
Kaohsiung, Taiwan
m1075503@mail.nuk.edu.tw

Wen-Yang Lin
*Dept. of Comp. Sci. and Info. Eng.*
*National University of Kaohsiung*
Kaohsiung, Taiwan
wylin@nuk.edu.tw

*Abstract*—**Spontaneous reporting systems (SRS) have been developed to collect adverse event records that contain personal demographics and sensitive information like drug indications and adverse reactions. The release of SRS data may disclose the privacy of the data provider. Unlike other microdata, very few anonymyization methods have been proposed to protect individual privacy while publishing SRS data. MS($k$, $\theta^*$)-bounding is the first privacy model for SRS data that considers multiple individual records, mutli-valued sensitive attributes, and rare events. PPMS($k$, $\theta^*$)-bounding then is proposed for solving cross-release attacks caused by the follow-up cases in the periodical SRS releasing scenario. A recent trend of microdata anonymization combines the traditional syntactic model and differential privacy, fusing the advantages of both models to yield a better privacy protection method. This paper proposes the PPMS-DP($k$, $\theta^*$, $\varepsilon$) framework, an enhancement of PPMS($k$, $\theta^*$)-bounding that embraces differential privacy to improve privacy protection of periodically released SRS data. We propose two anonymization algorithms conforming to the PPMS-DP($k$, $\theta^*$, $\varepsilon$) framework, PPMS-DP$_{num}$ and PPMS-DP$_{all}$. Experimental results on the FAERS datasets show that both PPMS-DP$_{num}$ and PPMS-DP$_{all}$ provide significantly better privacy protection than PPMS-($k$, $\theta^*$)-bounding without sacrificing data distortion and data utility.**

## I. Introduction

Since the Covid-19 pandemic ranged the globe in 2020, a skyrocketed amount of adverse events (AEs) related to Covid-19 vaccines or drugs has been reported to the spontaneous reporting system (SRS), like the USA FDA Adverse Drug Event Reporting System(FAERS) [6] and MedEffect Canada [3]. For example, the size of reports to US VAERS (Vaccine Adverse Event Reporting System) in 2021 is 623.19MB, nearly 15 times the size in 2020 (41.73MB) [7]. These adverse event records are valuable resources for researchers to analyze and detect actual adverse drug/vaccine event signals to monitor the safety of marketing drugs or vaccines. However, the SRS data collect patients' individual information, such as name, phone number, age, and gender, in addition to the drug information and reported indication. Hence, organizations or data holders need to consider privacy problems before releasing the records to specific researchers or the public.

One of the basic strategies to protect privacy is de-identification, i.e., removing explicit identifiers (*ED*) [13] that can be directly linked to the record owner, such as name and SSN. For example, the HIPPA privacy rule [1] requires the removal of 16 specific identifiers for publishing medical and health microdata. Even so, some attributes essential for signal detection are left, including quasi-identifiers (*QID*) such as age, gender, and sensitive attributes like drug information, drug indication, and adverse reaction. Researchers have shown that various privacy threats still exist for the de-identified medical and health microdata and proposed many privacy protection models and anonymization methods, such as $k$-anonymity [27], $l$-diversity [22], and $t$-closeness [18].

Lin *et al.* [20] first noticed some unique characteristics of SRS data that would paralyze previously proposed privacy protection methods. For example, most AE reports contain multivalued sensitive attributes, such as reaction and indication, meaning the anonymity models must consider several sensitive attributes while protecting a record. Besides, rare-event reports exist in the SRS data. Most anonymization methods would cause significant distortion for rare events and overlook AE signals related to rare events. Lin *et al.* [20] adjusted the mechanisms used in $k$-anonymity and $l$-diversity, proposing the MS($k$, $\theta^*$)-bounding model and the MS-anonymization algorithm. Later, Wang and Lin [28] observed the scenario of periodical publishing for SRS data. That is, the SRS data are released periodically, usually in a quarter, like FAERS. Besides, many follow-up cases containing complement or correction information of the original report are assigned the same CaseID for tracking purposes. The periodical publishing scenario along with follow-up CaseID opens the door for attackers to perform cross-release attacks. Wang and Lin identified three types of attacks, namely *Backward*-attack, *Forward*-attack, and *Latest*-attack, collectedly named *BFL*-attack, that will crack patients' privacy by joining different timestamped released tables through *QID* and CaseID. To protect periodical publishing SRS data from *BFL*-attack, they proposed the PPMS($k$, $\theta^*$)-bounding model and PPMS-Anonymization algorithm.

These models mentioned above are called syntactic anonymity methods [9], which require knowing the background information held by the attackers and aiming to defend against specific attacks, thus barely handling unknown types of attacks. Differential privacy, an emerging privacy protection model initially proposed by Dwork [10][11] for interactive query of databases, can protect privacy without assuming attackers' background knowledge. But differential privacy usually yields significant data distortion, making it unfeasible for medical and health microdata. This deficiency leads to an alternative trend by combing the syntactic anonymity model and differential privacy, such as ($k$, $\beta$)-SDGS [19], ($k$, $\varepsilon$)-anonymity [15], and IMDAV-DP [25][26]. Lin and Shen proposed MSDP($k$, $\theta^*$, $\varepsilon$) [21], an extension of MS($k$, $\theta^*$)-bounding by incorporating differential privacy to protect SRS data. However, MSDP($k$, $\theta^*$, $\varepsilon$) is designed for a single release of SRS data, not considering the privacy threat caused by follow-up cases in different releases.

In this paper, we propose a new differentially private protection method more suitable for protecting periodically released SRS data. The proposed model is PPMS-DP($k$, $\theta^*$, $\varepsilon$), a hybrid of differential privacy and PPMS($k$, $\theta^*$)-bounding. We also designed two anonymization methods conforming to the PPMS-DP($k$, $\theta^*$, $\varepsilon$) model: PPMS-DP$_{num}$ and PPMS-DP$_{all}$. A series of experiments conducted on the FAERS data show that PPMS-DP$_{num}$ and PPMS-DP$_{all}$ exhibit better privacy protection than PPMS++ [28], i.e., the best implementation of PPMS-Anonymization achieving PPMS($k$, $\theta^*$)-bounding, without sacrificing data utility for ADR detection.

The remainder of this paper is organized as follows. Section 2 introduces some background knowledge and related work. Section 3 presents our proposed two hybrid anonymization methods, PPMS-DP$_{num}$ and PPMS-DP$_{all}$. The empirical evaluation of our methods is described in Section 4. Finally, Section 5 summarizes the conclusion and presents some promising future work.

## II. Background Knowledge

### A. Privacy-Preserving Data Publishing

In the study of privacy-preserving data publishing of microdata, we can split attributes into four types [13]: explicit identifiers (*EID*s), quasi-identifiers (*QID*s), sensitive attributes (*SA*s), and non-sensitive attributes (*NSA*s). *EID*s are personal information that can identify the record owners, such as name, SSN, and an exact address. *QID*s are also personal information that cannot identify the record owners directly but can be linked with other data to raise the possibility of record identification, like gender, age, country, and job. *SA*s are sensitive information that record owners do not let others know. Most *SA*s denote health situations, medical treatment, or financial ability, such as drug indications, diseases, and salaries. *NSA*s refer to none of the above types of attributes, which would not cause privacy problems. These attributes usually are ignored in the course of data anonymization.

Sweeney first demonstrated how the attackers could identify the record owners through *QID* even without *EID*s. This attack is known as record linkage attack [27], causing privacy threats by joining two released tables. That is, the attackers re-identify a record's owner with external knowledge. Sweeney proposed the $k$-anonymity model that requires each group of records with the same *QID* value, also known as an equivalence class or *QID*-group, should contain at least $k$ records, limiting the possibility of successfully re-identifying the target record to at most $1/k$.

Although $k$-anonymity protects microdata from re-identification, it is inept at protecting the sensitive values of the target. This kind of attack focusing on *SA*s is known as attribute disclosure, also called attribute linkage attack. To prevent attribute disclosure, Machanavajjhala et al. [22] proposed the $l$-diversity model, a $k$-anonymity extension that requires each *QID*-group containing at least $l$ different sensitive values.

### B. Differential Privacy

Differential privacy [10] is an emerging privacy model with a rigorous theoretical foundation, aiming to prevent privacy disclosure from repeated query results of databases. The kernel concept of differential privacy is to maintain the query result not being affected by the existence or not of a specific record. Given a positive number $\varepsilon$, we say a randomized function $A$ satisfying $\varepsilon$-differential privacy, if for any two data sets $D_1$ and $D_2$ differing in at most one record for all possible generated result $S$ of $A$, we have

$$\frac{Pr[A(D_1) \in S]}{Pr[A(D_2) \in S]} \le e^{\varepsilon} \approx 1 + \varepsilon.$$

In brief, differential privacy ensures the difference between $A(D_1)$ and $A(D_2)$ is no more than $\varepsilon$. The smaller the privacy budget $\varepsilon$ is, the higher the requested privacy level.

The primary mechanism to achieve differential privacy is adding noise to the query result, which may cause the dilation of an attribute domain. The role of privacy budget $\varepsilon$ is to limit the dilation and avoid mass distortion. Generally, the noise is randomly generated according to the maximal difference of the query result from $D_1$ and $D_2$, called sensitivity. That is, given a function $f$: $D \rightarrow R^d$, the $L_1$-sensitivity of $f$ is

$$\Delta f = \max_{D_1, D_2} \| f(D_1) - f(D_2) \|_1$$

Dwork *et al.* [11] proposed a noise-adding mechanism, the Laplace mechanism, which adds random noise following Laplace distribution to numerical attributes or query results of microdata. They showed that the Laplace mechanism satisfies $\varepsilon$-differential privacy. Consider a numerical attribute $x$ in microdata. The amount of noise added to $x$ is,

$$A(x) = x + \mathrm{Lap}\left(\frac{\Delta f}{\varepsilon}\right),$$

where Lap($x$) denotes a Laplace distribution with zero mean and scale $x$.

The Laplace mechanism cannot add random noises to categorical attributes or non-numerical query results. McSherry and Talwar [23] proposed the exponential mechanism to determine the result of anonymized categorical attributes. The candidate results of a categorical *QID* attribute, which refer to a taxonomy tree, are nodes of the minimal subtree containing all values in the *QID*-group. Given a set of input data $D$ and the range $R$ of possible results, the probability of each anonymized result $r \in R$, $d \in D$ is

$$\exp\left(\frac{\varepsilon \times q(D, r)}{2 \times \Delta q}\right),$$

where $q(D, r)$ denotes the quality function to calculate the utility score between $D$ and $r$, and $\Delta q$ denotes the sensitivity of quality function $q$.

## III. Review of *BFL*-Attack and PPMS($k$, $\theta^*$)-Bounding

Lin *et al.* [20] presented several unique characteristics, such as rare events, multiple individual records, multivalued sensitive attributes, and missing values of SRS data that would paralyze traditional anonymization methods, like $k$-anonymity, $l$-diversity, etc. Another characteristic of SRS data is periodically releasing, which was first studied by Wang and Lin [28]. For example, USA FDA collects AE reports and publishes the FAERS data quarterly. They pointed out the problem of anonymizing each release independently, mainly caused by follow-up records in different-timestamped releases sharing identical CaseID. The attackers may use the CaseID to track a patient's history in different anonymized releases and perform cross-release attacks, namely *BFL*-attack. For illustration, consider a series of three anonymized tables shown in Table 1 that satisfy MS(4, 0.5)-bunding, which means every *QID*-group contains at least four records and the frequency of each sensitive value in the group is no more than 0.5.

**Backward-Attack.** Assume the attacker knows his neighbor Bob's *QID* is <Male, 37> and Bob's adverse event is in $R_3$. Then, *CI* for Bob is {1, 3, 7, 12}, where cases 1 and 3 occur in $R_1$ and $R_2$, while case 7 appears in $R_1$. Combining the information in $R_1$ and $R_2$, cases 1, 3, and 7 in $R_1$ fail to cover Bob's *QID*, so Bob's record is case 12 in $R_3$, revealing Bob has Diabetes and Flu.

**Forward-attack.** Assume the attacker knows his neighbor John's *QID* is <Male, 44> and John has an adverse event in $R_2$. Then, *CI* of John in $R_2$ is {1, 3, 8, 9}, of which cases 1 and 3 appear in $R_3$ as well. Since the age of cases 1 and 3 in $R_3$ fail to cover John's age, the attacker can exclude cases 1 and 3 from *CI* and concludes John has Diabetes.

**Latest-attack.** Assume the attacker knows his neighbor John's *QID* is <Male, 44> and Jane, female and age 30. Besides, the attacker knows Jane's adverse event first appears in $R_3$. Then, the matched CaseIDs of Jane in $R_3$ is {13, 14, 15, 16}. By checking previous releases $R_1$ and $R_2$, the attacker can exclude cases 13 and 14 and conclude Jane has Breast Cancer.

TABLE I. A SERIES OF THREE ANONYMIZED TABLES SATISFYING MS(4, 0.5)-BUNDING.

(a) Anonymized SRS table $R_1$

| CaseID | Gender | Age | Disease |
|---|---|---|---|
| 1 | Male | [20-30] | HIV, Fever |
| 2 | Male | [20-30] | Flu |
| 3 | Male | [20-30] | HIV |
| 4 | Male | [20-30] | Flu |
| 5 | Any | [30-35] | HIV |
| 6 | Any | [30-35] | HIV |
| 7 | Any | [30-35] | Diabetes, Flu |
| 8 | Any | [30-35] | Diabetes, Flu |

(b) Anonymized SRS table $R_2$

| CaseID | Gender | Age | Disease |
|---|---|---|---|
| 1 | Any | [30-45] | HIV, Fever |
| 3 | Any | [30-45] | HIV |
| 8 | Any | [30-45] | Diabetes, Flu |
| 9 | Any | [30-45] | Diabetes |
| 10 | Female | [20-45] | HIV |
| 11 | Female | [20-45] | Flu |
| 13 | Female | [20-45] | HIV, Flu |
| 14 | Female | [20-45] | Diabetes |

(c) Anonymized SRS table $R_3$

| CaseID | Gender | Age | Disease |
|---|---|---|---|
| 1 | Male | [20-40] | HIV, Fever |
| 3 | Male | [20-40] | HIV |
| 7 | Male | [20-40] | Diabetes, Flu |
| 12 | Male | [20-40] | Diabetes, Flu |
| 13 | Female | [20-45] | HIV, Flu |
| 14 | Female | [20-45] | Diabetes |
| 15 | Female | [20-45] | Breast Cancer |
| 16 | Female | [20-45] | Breast Cancer |

To prevent *BFL*-attack, Wang and Lin [28] proposed the PPMS($k$, $\theta^*$)-bounding privacy model and the PPMS-Anonymization algorithm.

**Definition 1. (PPMS($k$, $\theta^*$)-bounding)** [28] Let $S = \{s_1, s_2, \ldots, s_m\}$ be the set of all possible sensitive values in *SA* and $\theta^* = (\theta_1, \theta_2, \ldots, \theta_m)$ the probability thresholds for $S$, where $0 \leq \theta_j \leq 1$, for $1 \leq j \leq m$. A series of anonymized SRS data releases $R_1$, $R_2$, …, $R_n$ satisfy PPMS($k$, $\theta^*$)-bounding if

(1) The size of the candidate *QID*-group of each record $v$ in $R_i$ after excluding all vulnerable records leading to *BFL*-attack is no less than $k$, and

(2) The probability of inferring $v$ having any sensitive value $s_j \in S$ is no larger than $\theta_j$.

In practice, most sensitive values in SRS data are not so sensitive that they require high-level protection, such as common diseases like fever and headache. For this reason, PPMS($k$, $\theta^*$)-bounding allows a non-uniform setting of $\theta^*$; different sensitive values are specified to different confidence thresholds. The benefit is to reduce information loss, paying more attention to providing better protection for highly sensitive values such as HIV.

PPMS-Anonymization used two strategies, *QID*-bounding and *NC*-bounding, to prevent *BFL*-attack. *F*-attack occurs when the attacker can obtain a more detailed *QID* value from the current release to narrow the target range in some previously released table. Hence, *F*-attack can be prevented if the *QID* value of the target in the current release can cover all of its clones in previously released tables. This strategy is called *QID*-bounding.

**Definition 2 (*QID*-bounding).** Given a series of previously released tables $R_1$, …, $R_n$, the current table $R_i$ satisfies *QID*-bounding if the *QID* value of every record in $R_i$ covers that of its old cases in $R_1$, …, $R_n$.

Note that we cannot change the published dataset once a release is anonymized and published. *B*-attack and *L*-attack, unlike *F*-attack, crack the target's privacy in the current table via previously released tables that are unchangeable. Generalizing a record in the current table to a higher level is thus useless. Instead, new-CaseID records, as they have no corresponding old cases in the previous releases, will not be cracked by cross-release linkage and so can provide reliable protection in a *QID*-group. In this context, *NC*-bounding requires each *QID*-group in the currently released table containing at least $k$ new CaseID records to provide comparable performance as $k$-anonymity.

**Definition 3 (*NC*-bounding).** For each *QID*-group $g$ in a released table $R$ anonymized with $k$-anonymity, $R$ satisfies *NC*-bounding if $|g_{new}| \geq k$, where $g_{new}$ denotes the set of new records (with new CaseIDs) in $g$.

The main steps of PPMS-Anonymization are described as follows.

Step 1. Combine individual records with identical CaseID into a super record to solve multiple individual records and multivalued sensitive-attribute problems.

Step 2. Generalize each old CaseID record to cover its earliest anonymized clone as *QID*-bounding requires.

Step 3. Perform a clustering step to group super records into *QID*-groups, each of which satisfies PPMS($k$, $\theta^*$)-bounding and *NC*-bounding. Each *QID*-group $g$ grows by including an isolated record $r$ with minimal $\Delta IL(g, r) \times PR(g, r)$.

Step 4. Generalize super records in the same *QID*-group to become an equivalence class with the same *QID* value.

## IV. THE PROPOSED METHOD

This section presents the PPMS-DP method, an enhancement of PPMS-Anonymization that incorporates differential privacy to yield better protection for periodically released SRS data. We first introduce the basic concept of how to embrace differential privacy to the PPMS-Anonymization. Then we propose two algorithms based on the PPMS-DP framework, PPMS-$DP_{num}$ and PPMS-$DP_{all}$.

### A. Basic Concept

Since PPMS-Anonymization is a syntactic-based method that protects a record by hiding it in a crowd, the attacker can easily infer the *QID*-group where the record resides via the *QID* value of a target. To improve PPMS-Anonymization for better privacy protection without further assumption of external knowledge, we apply differential privacy to perturb some *QID* attributes to thwart the attacker's confidence in identifying the group in which the target resides identified by *QID* values. We choose to apply local differential privacy to each *QID*-group because it leads to less data distortion than global differential privacy and tends not to suppress rare events [8]. Furthermore, previous work has shown that achieving local differential privacy on *QID*-groups may provide sufficient privacy protection [15].

In light of the above discussion, we focus on revising the *QID*-grouping procedure of PPMS-Anonymization, applying different approaches for perturbing *QID*-group via differential privacy. Two options of differential privacy-based perturbation for *QID* attributes were considered, i.e., only disturbing numerical *QID* or all *QID* attributes, namely PPMS-$DP_{num}$ and PPMS-$DP_{all}$, respectively. To ease the discussion, we divide *QID* attributes into categorical ones $QID^C$ and numerical ones $QID^N$.

### B. Algorithm PPMS-$DP_{num}$

Algorithm PPMS-$DP_{num}$ is a variant of PPMS-Anonymization that adds noises only to numerical *QID* attributes, i.e., $QID^N$. To meet this strategy, we revise the kernel procedure of PPMS-Anonymization for dividing the records into *QID*-groups against *BFL*-attack and satisfying MS($k$, $\theta^*$)-bounding.

First, the revised grouping procedure divides the records in the current release $R_i$ into the set of new case records *NC* and the set of old case records *OC*. Then, perform the grouping function used in PPMS-Anonymization on *NC*, obtaining a set of *QID*-groups, each of which is composed of only new cases and a size of at least $k$. This result meets the *NC*-bounding strategy used in [28] to defend *BL*-attack.

Next, we assign the isolated new and all old cases into their most appropriate *QID*-group following the same criterion used in PPMS-Anonymization, i.e., minimizing $\Delta IL(g, r) \times PR(g, r)$, where $\Delta IL(g, r)$ represents the increase of information loss of a *QID*-group $g$ due to the inclusion of record $r$, while $PR(g, r)$ the privacy risk of *QID*-group $g$ caused by including $r$. The readers can refer to [28] for details of these formulas. We name this revised *QID*-grouping procedure New-Case-Core Grouping, shortly NCC-Grouping. Fig. 1 illustrates the concept of this procedure.

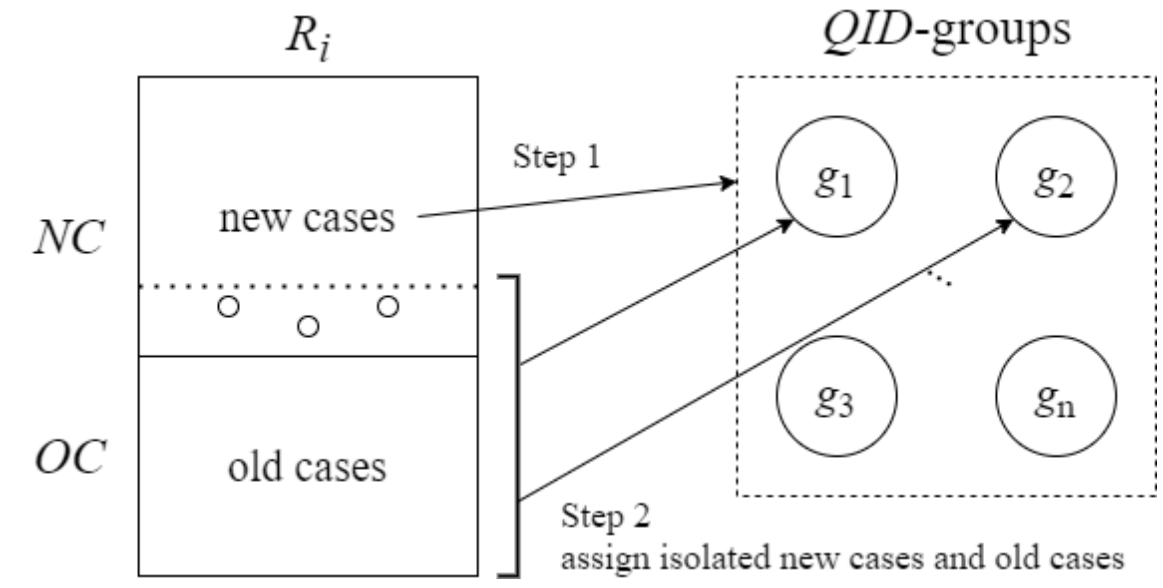

Fig. 1. An illustration of NCC-Grouping.

After the $QID^C$-covering procedure, we first generalize the $QID^C$ value within each *QID*-group. The reason for not considering $QID^N$ is that noise addition will later be applied to $QID^N$ values. The distortion of $QID^N$ attributes will be enlarged if we perform generalization on them. Then, an additional group merging procedure is applied to merge any *QID*-groups with identical $QID^C$ value to increase the diversity of $QID^N$ value for each *QID*-group so as to enlarge the sensitivity of the $QID^N$ attribute and improve the protection provided by applying the Laplace mechanism to $QID^N$. We name this procedure $QID^C$–Gen&Merging.

Finally, for each *QID*-group, we compute the local sensitivity of each $QID^N$ attribute within that group, denoted by $\Delta f(q)$, for $q \in QID^N$. Then, we perform the Laplace mechanism to add random noise to the attribute of each record $r$. That is,

$$v(q_r) = v(q_r) + Lap\left(\frac{\Delta f(q)}{\varepsilon}\right),$$

where $v(q_r)$ denotes the value of record $r$ for attribute $q$.

Fig. 2 describes the main steps of PPMS-$DP_{num}$, where we assume the life span of follow-up cases is at most $x$.

**Algorithm 1.** PPMS-$DP_{num}$
**Input:** The current dataset $D_i$, the previous anonymized releases $R = \{R_{i-x}, \cdots, R_{i-1}\}$, parameter $k$, confidence threshold $\theta^*$, and safety budget $\varepsilon$
**Output:** An anonymized dataset $R_i$

1. $D'$ ← the result of combining all records in $D_i$ with the same CaseID into a super record.;
2. $G$ ← {}; // The set of *QID*-groups.
3. $G$ ← NCC-Grouping($R$, $D'$, $k$, $\theta^*$);
4. $G$ ← $QID^C$-Covering($G$);
5. $G$ ← $QID^C$-Gen&Merging($G$);
6. $G$ ← LaplacePerturbation($G$, $\varepsilon$);
7. $R_i$ ← all records in $G$ are down posed to their original ones.
8. $D'$ ← all records recovered from super records in $G$;
9. **return** $R_i$;

Fig. 2. A summary of algorithm PPMS-$DP_{num}$

### C. Algorithm PPMS-$DP_{all}$

Algorithm PPMS-$DP_{all}$ adds noise to all *QID* attributes instead of $QID^N$ only. To apply differential privacy on $QID^C$, we modify some phases in PPMS-$DP_{num}$. First, PPMS-$DP_{all}$ does not need to consider *QID*-bounding. $QID^C$ generalization is replaced by $QID^C$ noise addition to preventing *F*-attack. Hence, we remove the $QID^C$-Covering function.

Second, we replace the $QID^C$-Gen&Merging function with a new function, VirtualGen&Merging. Unlike $QID^C$-Gen&Merging, we do not generalize the $QID^C$ value of all $QID^C$-groups, because these values will be sanitized later by an exponential mechanism. That is, we merge $QID^C$-groups if their "virtually" generalized $QID^C$ value is identical. This way can avoid unnecessary data distortion.

**Example 1.** Table II(a) shows a part of the clustering result, composed of two groups. Assume we can obtain the same categorical *QID* values {Gender = Any, Age = Young Adult} from group 1 and group 2 by generalization. Then both groups can merge due to having the same generalized categorical *QID* values. Table II(b) shows the final result of the merging.

TABLE II. AN EXAMPLE OF VIRTUAL GENERALIZING AND GROUP MERGING

(a) Two *QID* groups

| GID | Gender | Age | Weight |
|---|---|---|---|
| 1 | Male | Young Adult | 70 |
| 1 | Female | Young Adult | 69 |
| 1 | Male | Young Adult | 75 |
| 2 | Male | Young Adult | 65 |
| 2 | Female | Young Adult | 60 |
| 2 | Female | Young Adult | 55 |

(b) The resulting group

| GID | Gender | Age | Weight |
|---|---|---|---|
| 1 | Male | Young Adult | 70 |
| 1 | Female | Young Adult | 69 |
| 1 | Male | Young Adult | 75 |
| 1 | Male | Young Adult | 65 |
| 1 | Female | Young Adult | 60 |
| 1 | Female | Young Adult | 55 |

Third, we apply an extra noise addition function Exponential-Perturbation to $QID^C$, following the concept of exponential mechanism [25][26], which replaces the original value of a categorical *QID* attribute with a randomly chosen value from all possible results of that attribute. Consider a *QID*-group $g$ and one of its categorical attributes $C_i$. Let $dom(C_i, g)$ denote the set of values of $C_i$ in group $g$, $T$ the taxonomy tree of $C_i$, and $T^c$ be the minimal taxonomy tree that covers all attribute values in $dom(C_i, g)$. Then the set of candidate noise values for $C_i$ in group $g$, denoted by $\psi(C_i, g)$, includes all values in $dom(C_i, g)$ and their ancestors in $T^c$, that is,

$$\psi(C_i, g) = dom(C_i, g) \bigcup_{v \in dom(C_i, g)} anc(v, T^c)$$

where $anc(v, T^c)$ represents the set of ancestors of $v$ in tree $T^c$. Next, we define the quality $q(v, \psi)$ of a noise value $v$ in $\psi(C_i, g)$ as the total distortion (information lost) caused by replacing all values in $dom(C_i, g)$ by $v$.

$$q(v, \psi) = \sum_{u \in dom(C_i, g)} IL_c(u, v)$$

where $IL_c()$ is defined below; $\varphi_c(v)$ denotes the set of ancestors of value $v$ in $T^c$ including $v$ itself, i.e., $\varphi_c(v) = anc(v, T^c) \cup \{v\}$.

$$IL_c(u, v) = \frac{|\varphi_c(u) \cup \varphi_c(v)| - |\varphi_c(u) \cap \varphi_c(v)|}{|\varphi_c(u) \cup \varphi_c(v)|}$$

The sensitivity $\Delta q$ of quality function $q$ can be defined as the difference between the maximum and minimum of $IL_c(u, v)$.

$$\Delta q = \max_{u \in dom(C_i, g), v \in \psi(C_i, g)} IL_c(u, v) - \min_{u \in dom(C_i, g), v \in \psi(C_i, g)} IL_c(u, v)$$

Then a noise value is randomly chosen following the exponential probability distribution $\exp(\varepsilon \times -q(v, \psi)/2\Delta q)$.

In short, the proposed noise perturbation for categorical attributes fuses generalization and exponential mechanism.

**Example 2.** Consider the Age taxonomy tree in Fig. 3 and the group in Table III. Then $dom(\text{Age}, g)$ = {Child, In-school, Adolescent}. The $T^c$ of $dom(\text{Age}, g)$ is shown in Fig. 4. We have $\psi(\text{Age}, g)$ = {Child, In-school, Adolescent, Non-adult}. To perform the proposed exponential mechanism to sanitize the Age attribute of group $g$, we need to compute $q(v, \psi)$ for every $v$ in $\psi(\text{Age}, g)$. For example,

$$q(\text{Child}, \psi) = IL_c(\text{Child, In-school}) + IL_c(\text{Child, Adolescent}) + IL_c(\text{Child, Non-adult})$$
$$= (3 - 2)/3 + (3 - 1)/3 + (2 - 1)/2 = 1.5$$

In the same way, we compute $IL_c$ for all other values and obtain

$$\max_{u \in dom(C_i, g), v \in \psi(C_i, g)} IL_c(u, v) = 0.75$$

$$\min_{u \in dom(C_i, g), v \in \psi(C_i, g)} IL_c(u, v) = 0.5$$

Hence, $\Delta q = 0.25$. Assume $\varepsilon = 0.1$. The probability of replacing the Age value of group $g$ with "Child" according to our designed exponential mechanism is

$$\exp\left(\frac{0.1 \times -1.5}{2 \times 0.5}\right) = 0.86$$

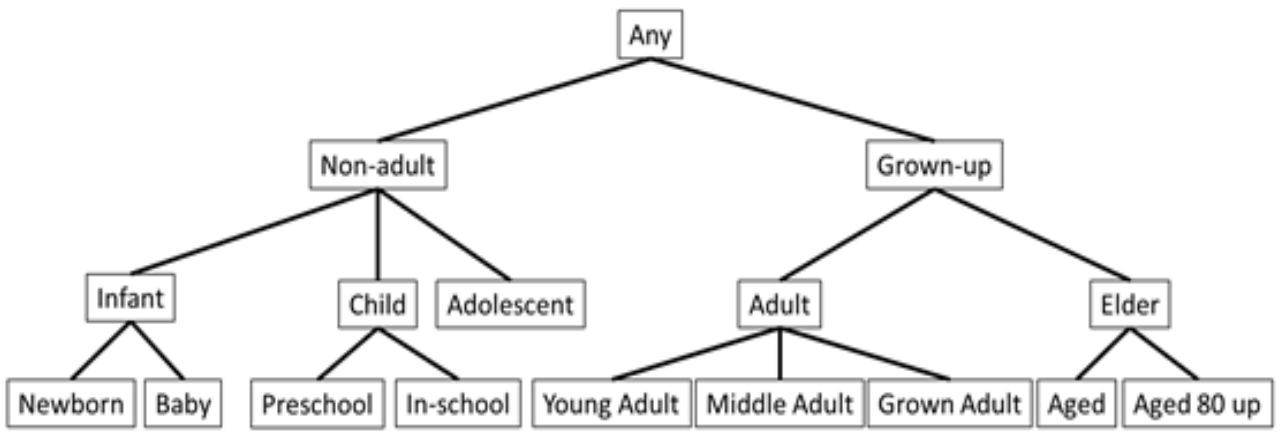

Fig. 3. The taxonomy tree for Age.

TABLE III. A *QID*-GROUP AFTER CLUSTERING.

| GID | Gender | Age | Weight |
|---|---|---|---|
| 1 | Female | Child | 30 |
| 1 | Female | In-school | 35 |
| 1 | Female | Adolescent | 45 |

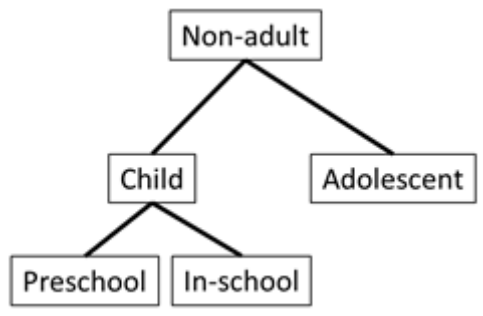

Fig. 4. The corresponding $T^c$ of the *QID*-group in Table III.

Finally, we adopted a new information loss ($IL^+$), which is more feasible for measuring data distortion caused by noise addition perturbation and generalization. The definition of $IL^+(g)$ is as follows.

$$\sum_{j=1}^{|g|} \left( \sum_{i=1}^{m} \frac{|\alpha(r'_j, N_i) - \alpha(r_j, N_i)|}{max(N_i) - min(N_i)} + \sum_{i=1}^{n} \frac{|\varphi(\alpha(r'_j, C_i)) \cup \varphi(\alpha(r_j, C_i))| - |\varphi(\alpha(r'_j, C_i)) \cap \varphi(\alpha(r_j, C_i))|}{|\varphi(\alpha(r'_j, C_i)) \cup \varphi(\alpha(r_j, C_i))|} \right)$$

where $\max(N_i)$ and $\min(N_i)$ denote the maximal and minimal value of numerical attribute $N_i$, function $\alpha(r_j, N_i)$ ($\alpha(r_j, C_i)$) represents the value of record $r_j$ in $N_i$ ($C_i$), $r'_j$ is the anonymized $r_j$, $|g|$ denotes the size of group $g$, and $\varphi(\alpha(r_j, C_i))$ denotes the set of ancestors of $\alpha(r_j, C_i)$ along with $\alpha(r_j, C_i)$ in taxonomy tree $T_i$ of $C_i$. The main steps of PPMS-DP$_{\text{all}}$ are described in Fig. 5.

**Algorithm 2.** PPMS-DP$_{all}$
**Input:** The current dataset $D_i$, the previous anonymized releases $R = \{R_{i-x}, \cdots, R_{i-1}\}$, parameter $k$, confidence threshold $\theta^*$, and privacy budget $\varepsilon$
**Output:** An anonymized dataset $R_i$

1. $D'$ ← the result of combining all records in $D_i$ with the same CaseID into a super record.;
2. $G$ ← {}; // Initialize the set of *QID*-groups.
3. $G$ ← NCC-Grouping($R$, $D'$, $k$, $\theta^*$);
4. $G$ ← VirtualGen&Merging($G$);
5. $G$ ← LaplacePerturbation($G$, $QID^N$, $\varepsilon$); // for $QID^N$
6. $G$ ← ExponentialPerturbation($G$, $QID^C$, $\varepsilon$); // for $QID^C$
7. $R_i$ ← all records in $G$ are down posed to their original ones.
8. $D'$ ← all records recovered from super records in $G$;
9. **return** $R_i$;

Fig. 5. A summary of algorithm PPMS-DP$_{all}$.

## V. Empirical Evaluation

### A. Experimental Design

To evaluate the performance of our proposed methods, we considered four aspects of measurement, including information loss, record disclosure risk, attribute disclosure risk, and signal bias. We used the FAERS dataset from 2004Q1 to 2011Q4. To simplify the complexity, we chose the following attributes, drug name (DRUGNAME), CaseID (CSAEID), age (AGE), gender (GNDR_COD), weight (WT), reaction (PT), and drug indication (INDI_PT) by joining the related tables through PRIMARYID. The drug names were standardized following the procedure in [29]. All records containing missing values in any attribute were excluded. For each record, we used {Gender, Age} as categorical *QID*s and {Weight} as numerical *QID*. Sensitive values include drug indication (INDI_PT) and drug reaction (PT); both are multivalued attributes. Besides, we simulated the *BFL*-attack on released tables by linking records with the same CaseID to evaluate the privacy risk yielded by each anonymization method.

To evaluate the data distortion caused by anonymization, we used Normalized Information Loss (*NIL*) to calculate the average differences for each record before and after anonymization. We adopted the information loss used in [28].

$$NIL(D') = \frac{\sum_{g \in D'} IL^*(g)}{|QID| \times |g|},$$

where $D'$ represents the anonymized version of dataset $D$, $g$ denotes a group in $D'$, $|QID|$ is the cardinality of $QID$, and $|g|$ is the number of records in group $g$. And, $IL^*$ is a generalized version of $IL^+$ on accounting for generalized numerical values. Let $U$ and $L$ denote the lower and upper bounds of a generalized value of $r_j$ at attribute $N_i$. $IL^*$ replaces the difference $\alpha(r'_j, N_i) - \alpha(r_j, N_i)$ used in $IL^+$ as

$$\alpha(r'_j, N_i) - \alpha(r_j, N_i) = \frac{\int_L^U |x - \alpha(r_j, N_i)| dx}{U - L}.$$

The two types of privacy disclosure risk, record identify and attribute disclosure risk, are measured by *RR* and *AR*. The *RR*, proposed in [25], calculates the total risk on record linkage disclosure as

$$RR(D') = \frac{\sum_{r' \in D'} Pr(r')}{|D'|},$$

where $r'$ denotes the anonymized version of record $r$, $D'$ the anonymized dataset of $D$, and $Pr(\cdot)$ is the probability of successfully identifying the target's record, defined as

$$Pr(r') = \begin{cases} 0 & \text{if } r' \notin G \\ \frac{1}{|G|} & \text{if } r' \in G' \end{cases}$$

where $G$ is the set of records in $D'$ with the minimum difference from $r$. That is, $G$ represents the set of possible anonymized versions of $r$, each of which is very similar to $r$.

There have been some different measurements of attribute disclosure risk, for example, *DSR* [28] and *AR* [21]. To achieve a more reasonable measure, we propose a revised version of *AR*, namely $AR_{\mathrm{rev}}$, which calculates the probability that the attacker can successfully infer any anonymized record's sensitive values. $AR_{\mathrm{rev}}$ is defined as below, an average over all the *Ar* for all records in $D'$.

$$AR_{\mathrm{rev}}(D') = \frac{\sum_{r \in D'} Ar(r')}{|D'|}.$$

For this purpose, we have to measure $Ar(r')$, the probability of successfully inferring any sensitive value of $r'$, which is defined as follows.

$$Ar(r') = \begin{cases} 0, & \text{if } r' \notin G \\ \frac{\sum_{s \in S_G} \max\{1/|G|, Pr_G(s)\}}{|S_G|}, & \text{if } r' \in G' \end{cases}$$

where $Pr_G(s)$ denotes the frequency of sensitive value $s$ in group $G$, and $S_G$ is the set of sensitive values in $G$. This function fuses two concepts, record identity and attribute disclosure. The sensitive value is also explored if an attacker can infer the target's record. This yields the probability $1/|G|$, similar to *RR*. Besides, the attacker also can infer the sensitive value $s$ with probability $Pr_G(s)$. So the resulting probability is $\max\{1/|G|, Pr_G(s)\}$.

The data utility measures how reliable the results analyzed from anonymized data are. In the context of ADR signal detection, we considered the following severe adverse drug reaction caused by AVANDIA, calculating the signal differences between the original dataset and the anonymized version.

$$\text{AVANDIA, age} > 18 \rightarrow CERECBROVASCULAR\ ACCIDENT$$

There have been several methods for measuring the strength of ADR signals [24]. In this study, we adopted the most common PRR [12], defined below.

$$\text{PRR} = \frac{a/(a+b)}{c/(c+d)},$$

where $a$, $b$, $c$, $d$ are the observed occurrences in the contingency table in Table II.

TABLE IV. Contingency table for ADR signal.

| | Reaction *R* | Other reaction*s* |
|---|---|---|
| **Drug *D*** | *a* | *b* |
| **Other drugs** | *c* | *d* |

Three parameters would affect the performance of each method. They are the size of anonymous group $k$ ($k$ = 5, 10, 15, 20), confidence bounding ($\theta^*$ = 0.2, 0.4, level-wise), and privacy budget $\varepsilon$ ($\varepsilon$ = 0.1, 1, 10). The level-wise setting for $\theta^*$ followed the concept in [20]. We divided all sensitive values into two types, sensitive and non-sensitive. Sensitive values include most information about sexually transmitted diseases, such as HIV and related medicine. Due to similar results observed, we omit $k$ = 10, 15, and $\varepsilon$ = 1. Besides, the performance resulting from $\theta^*$ = level-wise is very similar to $\theta^*$ = 0.4. We also skip the level-wise setting.

### B. *Results on NIL*

In this section, we show the results of data distortion measured by *NIL* with confidence bounding $\theta^*$ = 0.2, 0.4 and privacy budget $\varepsilon$ = 0.1 and 10. Note parameter $\varepsilon$ is not applicable for PPMS++, which is applied only on PPMS-$DP_{num}$ and PPMS-$DP_{all}$.

As shown in Fig. 6, we observe that PPMS-$DP_{num}$ and PPMS-$DP_{all}$ produce much more information loss than PPMS-anonymization. However, the difference decreases as the privacy budget $\varepsilon$ increases. Besides, the data distortion caused by PPMS-$DP_{num}$ and PPMS-$DP_{all}$ is nearly not affected by $k$, even though different $k$'s would lead to different clustering results.

Although the difference in *NIL* for PPMS-$DP_{num}$ and PPMS-$DP_{all}$ is not significant, in general, PPMS-$DP_{all}$ yields less information loss than that of PPMS-$DP_{num}$. This is because PPMS-$DP_{all}$ adopts the proposed fusion of generalization and exponential mechanism, which maintains better semantic information and prevents a more considerable distortion caused by the more general generalization performed by PPMS-$DP_{num}$.

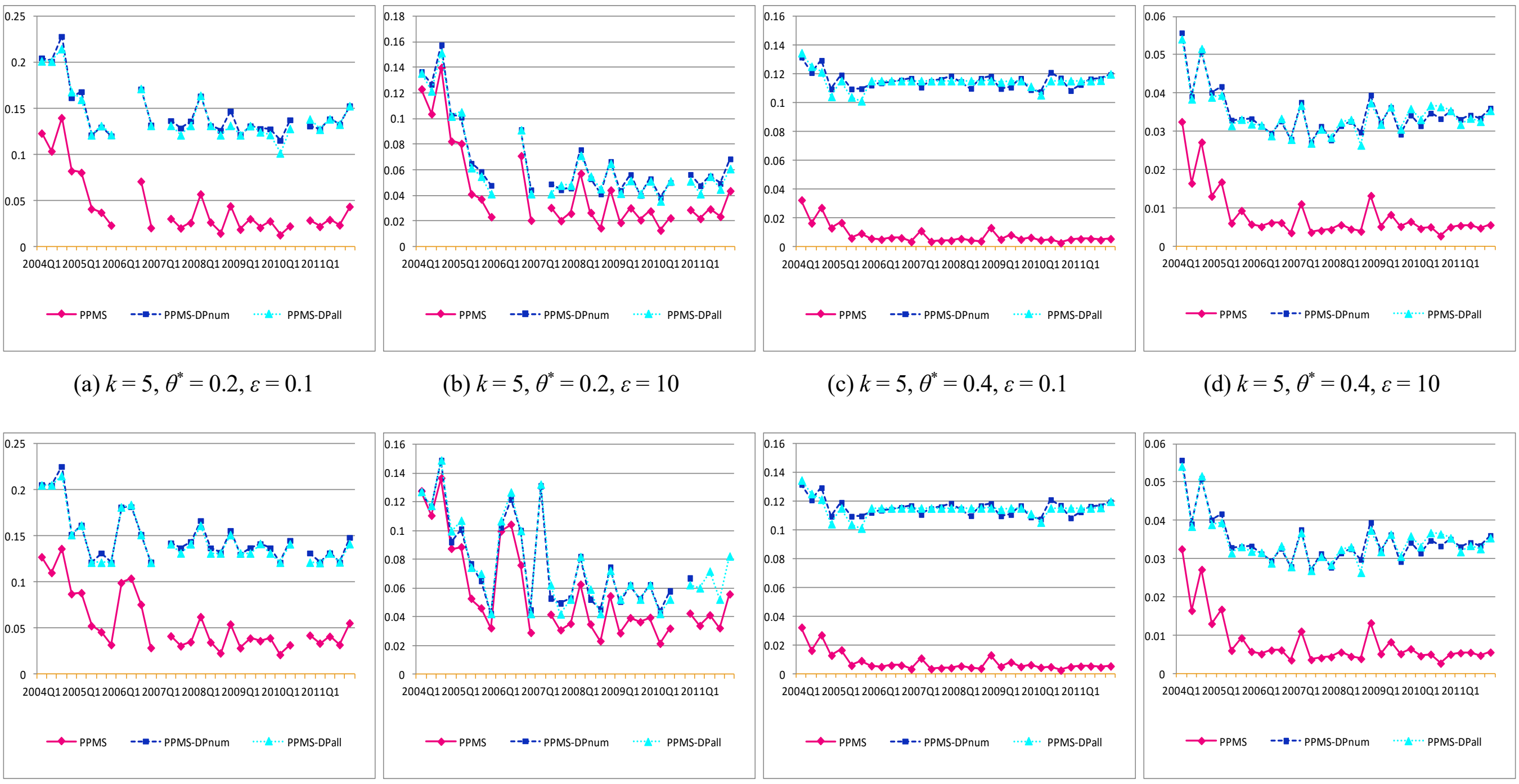

(a) $k = 5$, $\theta^* = 0.2$, $\varepsilon = 0.1$ (b) $k = 5$, $\theta^* = 0.2$, $\varepsilon = 10$ (c) $k = 5$, $\theta^* = 0.4$, $\varepsilon = 0.1$ (d) $k = 5$, $\theta^* = 0.4$, $\varepsilon = 10$

(e) $k = 20$, $\theta^* = 0.2$, $\varepsilon = 0.1$ (f) $k = 20$, $\theta^* = 0.2$, $\varepsilon = 10$ (g) $k = 20$, $\theta^* = 0.4$, $\varepsilon = 0.1$ (h) $k = 20$, $\theta^* = 0.4$, $\varepsilon = 10$

Fig. 6. Comparison on *NIL*s ($k = 20$, $\theta^* = 0.2, 0.4$, $\varepsilon = 0.1, 1, 10$).

### C. *Results on RR and AR*

Next, we present the results of privacy protection measured by *RR* and *AR*. According to Figs. 7 and 8, we observe that PPMS++ is significantly worse than PPMS-$DP_{num}$ and PPMS-$DP_{all}$, either on the results of *RR* or *AR*. In the worst case, PPMS++ generates more than 3% of record risk and attribute risk. On the other hand, either *RR* or *AR* yielded by PPMS-$DP_{num}$ and PPMS-$DP_{all}$ is less than 0.6%, even with a larger privacy budget ($\varepsilon$ = 10). PPMS-$DP_{num}$ and PPMS-$DP_{all}$ exhibit similar results on *AR* and *RR*, but PPMS-$DP_{all}$ performs slightly better than PPMS-$DP_{num}$.

### D. *Influence on ADR Signal*

Finally, we present the result of signal bias with $k$ = 5, 20, $\varepsilon$ = 0.1, 10, and $\theta^*$ = 0.2, 0.4. From Fig. 9, we observe that the results between PPMS-$DP_{num}$ and PPMS++ are overlapping in most of the time. Only when $k$ is large ($k$ = 20), we can observe the difference. It may be because both algorithms use a similar bounding strategy caused by similar clustering results. PPMS-$DP_{all}$ outperforms PPMS-$DP_{num}$ and PPMS++ in nearly all situations.

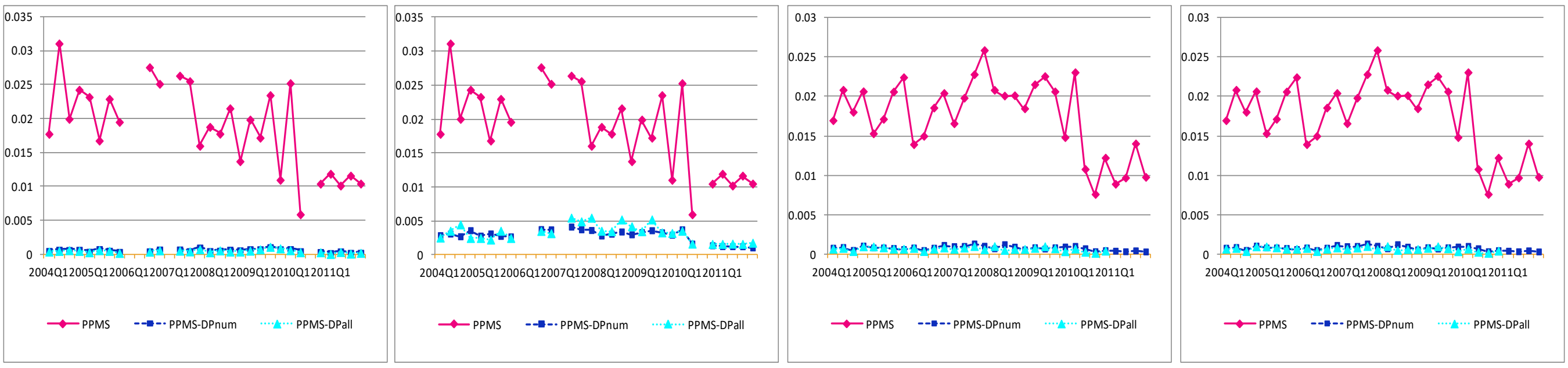

(a) $k = 5$, $\theta^* = 0.2$, $\varepsilon = 0.1$ (b) $k = 5$, $\theta^* = 0.2$, $\varepsilon = 10$ (c) $k = 5$, $\theta^* = 0.4$, $\varepsilon = 0.1$ (d) $k = 5$, $\theta^* = 0.4$, $\varepsilon = 10$

Fig. 7. Comparison on *RR*s ($k = 5$, $\theta^* = 0.4$, $\varepsilon = 0.1, 1, 10$).

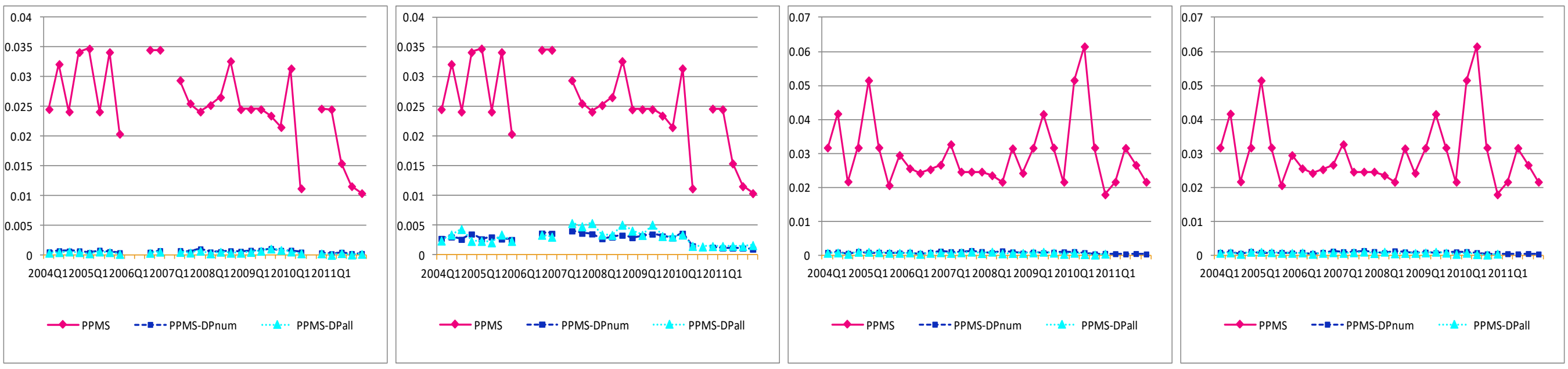

(a) $k = 5$, $\theta^* = 0.2$, $\varepsilon = 0.1$ (b) $k = 5$, $\theta^* = 0.2$, $\varepsilon = 10$ (c) $k = 5$, $\theta^* = 0.4$, $\varepsilon = 0.1$ (d) $k = 5$, $\theta^* = 0.4$, $\varepsilon = 10$

Fig. 8. Comparison on *AR*s ($k = 5$, $\theta^* = 0.2, 0.4$, $\varepsilon = 0.1, 1, 10$).

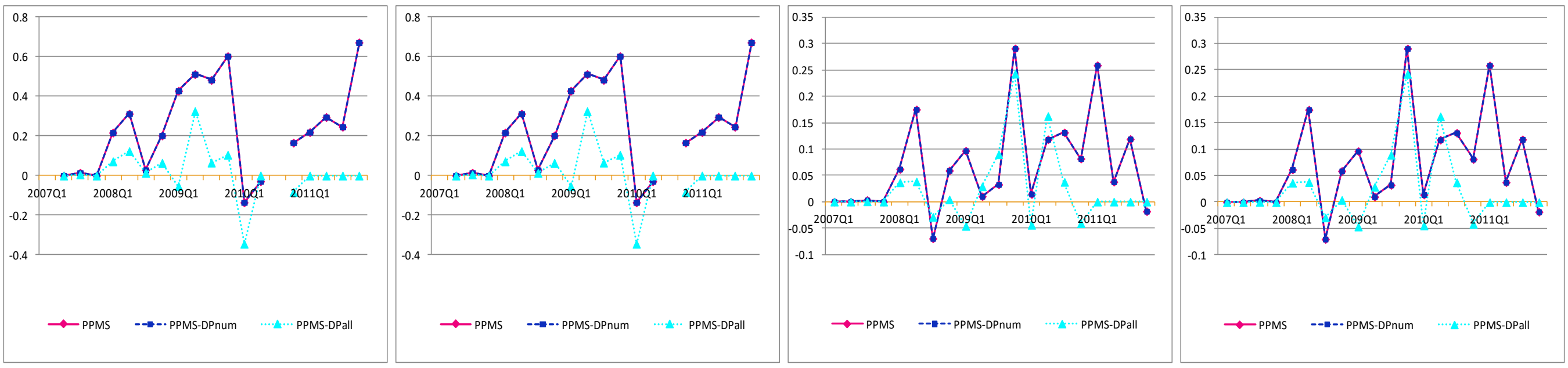

(a) $k = 5$, $\theta^* = 0.2$, $\varepsilon = 0.1$ (b) $k = 5$, $\theta^* = 0.2$, $\varepsilon = 10$ (c) $k = 5$, $\theta^* = 0.4$, $\varepsilon = 0.1$ (d) $k = 5$, $\theta^* = 0.4$, $\varepsilon = 10$

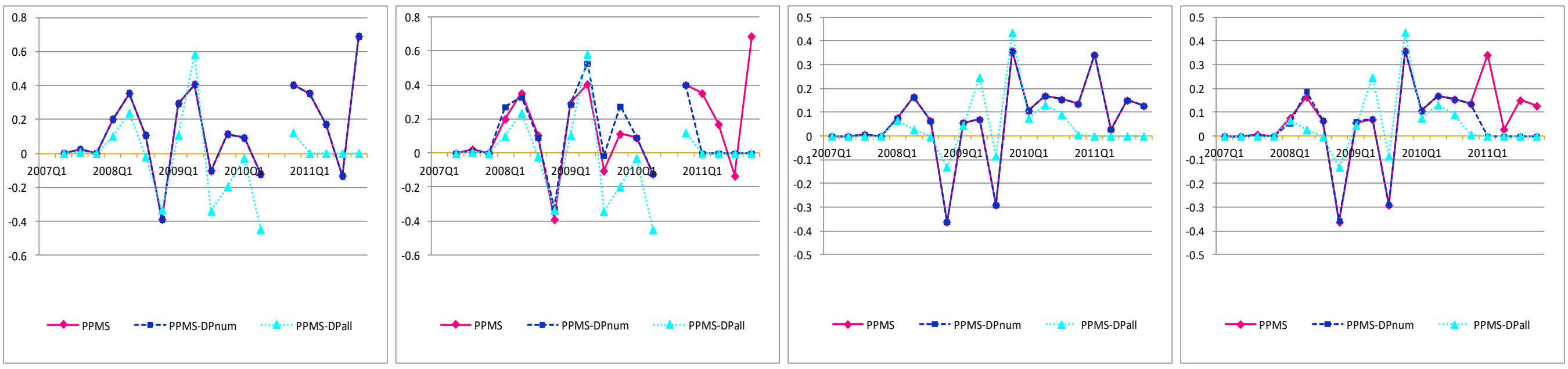

(e) $k = 20$, $\theta^* = 0.2$, $\varepsilon = 0.1$ (f) $k = 20$, $\theta^* = 0.2$, $\varepsilon = 10$ (g) $k = 20$, $\theta^* = 0.4$, $\varepsilon = 0.1$ (h) $k = 20$, $\theta^* = 0.4$, $\varepsilon = 10$

Fig. 9. Comparison on ADR signal bias

## *E. Evaluation on Medical Background Knowledge*

As differential privacy is well known for providing more robust protection without knowing the background knowledge held by the attacker, we conducted another experiment to examine whether our differential privacy fusing algorithms can prevent further attacks by utilizing some common medical knowledge. For this purpose, diseases related to specific gender and age group were considered additional background knowledge to the attacker. Female-related diseases include Breast Cancer, Cervitivis, and Polycystic. Male-related diseases include Prostate Cancer and Hernia. Also, elderly-related diseases reveal extra age knowledge, including Chronic Obstructive Pulmonary Disease (COPD) and Alzheimer's disease. In this experiment, we generated a new dataset combining 2009Q2, 2010Q1, and 2010Q3.

Fig. 10 shows the *RR* bias between the results with and without extra background knowledge. Our two methods yield nearly zero bias even with extra medical knowledge, but PPMS++ exhibits additional privacy risk. Similar results are observed for *AR* bias in Fig. 11.

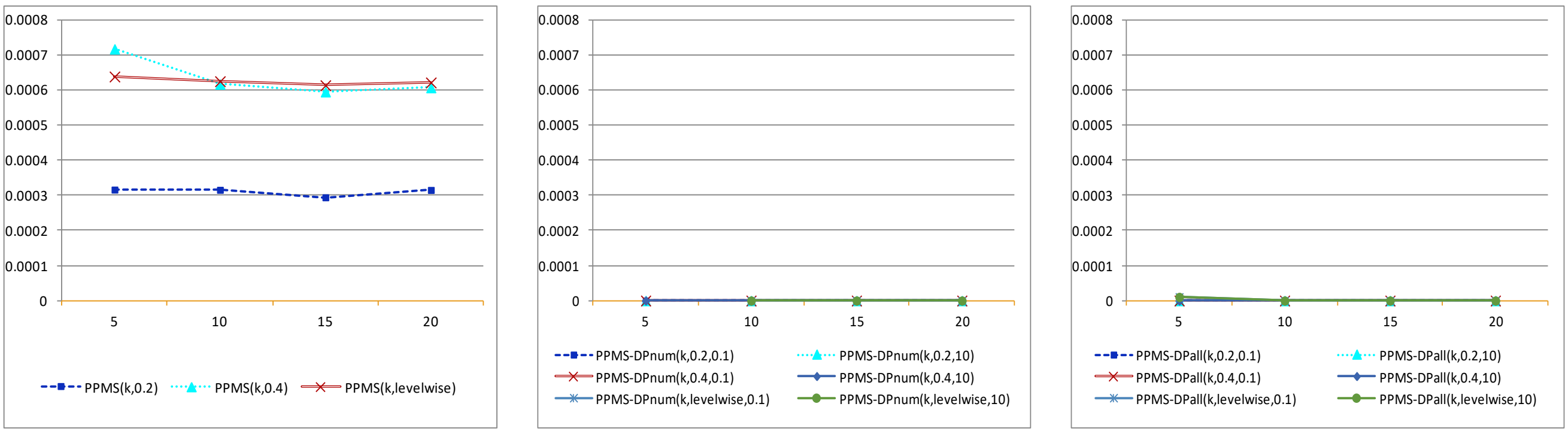

(a) PPMS($k$, $\theta^*$) (b) PPMS-DP$_{num}$($k$, $\theta^*$, $\varepsilon$) (c) PPMS-DP$_{all}$($k$, $\theta^*$, $\varepsilon$)

Fig. 10. Comparison of *RR* bias considering health background knowledge.

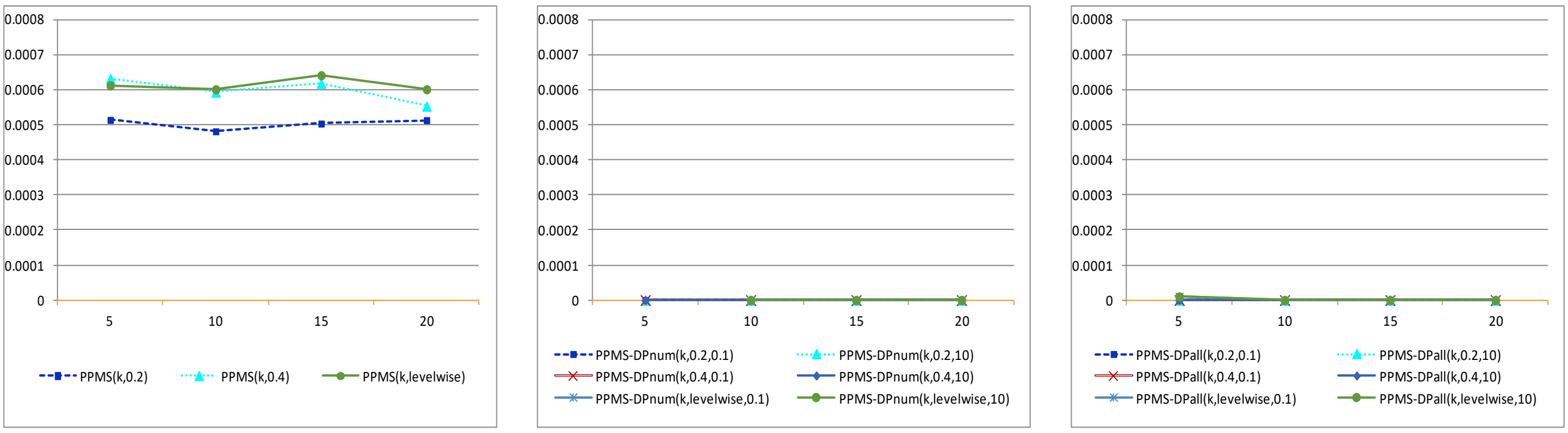

(a) PPMS($k$, $\theta^*$) (b) PPMS-DP$_{num}$($k$, $\theta^*$, $\varepsilon$) (c) PPMS-DP$_{all}$($k$, $\theta^*$, $\varepsilon$)

Fig. 11. Comparison of *AR* bias considering health background knowledge.

## VI. Conclusion

Hybrid anonymization methods that combine the syntactic model and differential privacy have become a new research trend in privacy protection of microdata. However, very few works have considered anonymizing SRS data in a periodical releasing scenario. Considering the *BFL*-attack noticed by Wang and Lin, in this paper, we have proposed a new privacy framework embracing differential privacy, called PPMS-DP($k$, $\theta^*$, $\varepsilon$). This framework enhances PPMS($k$, $\theta^*$)-bounding by leveraging the power of differential privacy to provide better privacy protection against *BFL*-attack in the periodical publishing scenario of SRS data. Based on the PPMS-DP($k$, $\theta^*$, $\varepsilon$) framework, we have also developed two algorithms, PPMS-DP$_{num}$ and PPMS-DP$_{all}$, to anonymize a new release of SRS data. The main difference between these two algorithms is that PPMS-DP$_{num}$ adds differential noise only to the numerical *QID* values and applies generalization on categorical *QID* values, while PPMS-DP$_{num}$ performs differential perturbation to all *QID* values. To evaluate our proposed methods, we have conducted a series of experiments using the well-known FAERS data. We have considered four performance measures, including information loss, record risk, attribute risk, and impact on ADR signal. Results show that PPMS-DP$_{num}$ and PPMS-DP$_{all}$ provide significantly better privacy protection than PPMS-Anonymization without sacrificing data utility for signal strength. Noteworthily, PPMS-DP$_{all}$ suffers lesser privacy threat from *BFL*-attack than PPMS-DP$_{num}$ and induces less information loss. PPMS-DP$_{all}$, which adopts a clever way to fuse differential perturbation to all *QID* values, is more suitable for the periodical released publishing of SRS data.

Another critical characteristic of SRS data is containing a lot amount of missing values. Unfortunately, most contemporary anonymization approaches overlook the impact of missing values [16]. We will extend the proposed methods to manage missing values.

## Acknowledgment

This work was supported by the Ministry of Science and Technology of Taiwan under grant no. MOST108-2221-E-390-016.